\documentclass[aps,prl,amsmath,amssymb,onecolumn,nofootinbib,showpacs,tightenlines,12pt]{revtex4}
\usepackage{epsfig,graphicx}
\usepackage{bm}
\newcommand{\be}{\begin{equation}}
\newcommand{\ee}{\end{equation}}
\newcommand{\bea}{\begin{eqnarray}}
\newcommand{\eea}{\end{eqnarray}}

\begin{document}
\title{ Some aspects of the  Liouville equation in mathematical physics and statistical mechanics}
\author{Ali Reza Khatoon Abadi}
\affiliation{Department of Mathematics, Islamic azad
university,Tehran's West branch,Tehran,IRAN}

\author{Ali morovvatpoor}
\affiliation{Science faculty ,department of mathematics,Payame noor
university }
\author{Mohammad Mehrpooya}
\affiliation{Department of Mathematics, Faculty of Science, Zabol
University, Zabol, Islamic Republic of Iran.}
\author{H.R.Rezazadeh}
\affiliation{Department of Mathematics,Islamic Azad university,KARAJ
branch,Karaj,IRAN} \email{h-rezazadeh@kiau.ac.ir}
\author{F.Golgoii}
 \affiliation{Department of Mathematics,Islamic Azad university,KARAJ branch,Karaj,IRAN}
\begin{abstract}
 This paper presents  some  mathematical aspects of
 Classical Liouville theorem and we have noted
 some mathematical theorems about its initial value problem.
 Furthermore, we have implied on the formal frame work of Stochastic
 Liouville equation (SLE).
 \end{abstract} \maketitle
\section{\label{sec:level1}Introduction}
This short essay has two purposes:(I) A rapid review of classical
and stochastic quantum  Liouville equation and proof of that and
some use of Liouville equation in statistical mechanics  and (II) to
present a short solution for a simple solvable model.The Liouville
equation is valid for both equilibrium and non-equilibrium
systems.It is imperative to the proof of fluctuation theorem from
which the second law of thermodynamics can be derived and also it is
the key component of derivation of Green Kubo relation \cite{1}for
linear transport coefficients such as shear viscosity,thermal
conductivity or electrical conductivity.
\\
The dynamics of such composite systems is governed by a
quantum-classical Liouville equation for either the density matrix
or dynamical variables which are operators in the Hilbert space of
the quantum subsystem and functions of classical phase space
variables of the classical enviroment.For more applications in
quantum mechanics refer \cite{2}.
\section{i.Description of classical Liouville equation and its proof}
The Liouville equation is a cardinal equation of statistical
mechanics.This equation depicts the development of phase space
distribution function for the conservative Hamiltonian system and
also supplies a complete description of the system both at
equilibrium and away from equilibrium. On the other hand Liouville
equation is a persistent for the flux and meditate the phase space
of micro-canonical ensemble of a N-particle system(NVE). Let $
\rho(q^{(N)},p^{(N)},t)$ refer the phase space density. If we
contemplate a volume element $ dp^{(N)} dq^{(N)}$ ,then
$\rho(q^{(N)},p^{(N)},t)dp^{(N)}dq^{(N)}$ acquire the number of
distinctive points in volume element.

It is a typical form of equation of motion of $
\rho(q^{(N)},p^{(N)},t).$

The number of system in the ensemble preserved the abidance equation
for development of phase space density assumed as:

\begin{eqnarray}
\nonumber\frac{\partial\rho}{\partial t} = -\nabla .
\overrightarrow{J}\\\nonumber\frac{\partial\rho}{\partial t}= -\nabla  .\rho\nu\\
\frac{\partial\rho}{\partial t}=
-\sum^{N}_{i=1,\alpha=1,3}[\frac{\partial}{\partial q_{i
\alpha}}(\rho \dot{q_{i \alpha}})+\frac{\partial}{\partial p_{i
\alpha}}(\rho \dot{p_{i \alpha}})]
\end{eqnarray}
implementing chain rule one obtains
\begin{eqnarray}
\frac{\partial \rho}{\partial t}= -\sum^{N}_{i=1,\alpha=1,3}
(\frac{\partial \rho}{\partial q_{i\alpha}} \dot{q_{i\alpha}} +
\frac{\partial \rho}{\partial p_{i\alpha}} \dot{p_{i\alpha}}) -
\rho( \sum^{N}_{i=1 , \alpha= 1,3}(\frac{\partial
\dot{q_{i\alpha}}}{\partial q_{i\alpha}} + \frac{\partial
\dot{p_{i\alpha}}}{\partial p_{i\alpha}}))
\end{eqnarray}
\\
 In  a Hamiltonian conservative system, the energy is kept up as a function of time and
the time derivatives are measured by Hamiltonian equations of motion
of classical mechanics.
\\
They are assumed by a set of coupled first
order partial differential equations ,
\begin{eqnarray}
\dot{q_{i \alpha}}= \frac{\partial H}{\partial p_{i \alpha}}\\
\dot{p_{i \alpha}}= - \frac{\partial H}{\partial q_{i
\alpha}}\\\nonumber i=1,2,...,N\\\nonumber \alpha=1,2,3
\end{eqnarray}
where   $q_{i\alpha}$   is the $\alpha-th$ component of the
situation,$q_{i}$ of the $ i-th$ particle.
\\
If we use Hamilton's equation of motion we acquire:
\begin{eqnarray}
\frac{\partial \dot{q_{i \alpha}}}{\partial q_{i \alpha}}=
\frac{\partial^2 H}{\partial q_{i \alpha} \partial
p_{i \alpha}}\\
\frac{\partial \dot{p_{i \alpha}}}{\partial P_{i \alpha}}= -
\frac{\partial^2 H}{\partial p_{i \alpha} \partial q_{i \alpha}}
\end{eqnarray}
thus the last term in equation (4) is indistinguishably zero leaving
us with
\begin{eqnarray}
\nonumber \frac{\partial \rho}{\partial t}=
-\sum_{i,\alpha}[\frac{\partial \rho}{\partial q_{i \alpha}}
\dot{q_{i \alpha}} + \frac{\partial
\rho}{\partial p_{i \alpha}} \dot{\rho_{i \alpha}}]\\
\frac{\partial \rho}{\partial t}= - \sum_{i,\alpha}[\frac{\partial
\rho}{\partial q_{i \alpha}}\frac{\partial H}{\partial \rho_{i
\alpha}} + \frac{\partial \rho}{\partial p_{i \alpha}}\frac{\partial
H}{\partial q_{i \alpha}}]
\end{eqnarray}
The last equation is the classical Liouville equation. Some of
scientists can use the classical Liouville equation symbolically :
\begin{eqnarray}
\frac{\partial \rho}{\partial t} = [H,\rho]
\end{eqnarray}
$Proof : $ consider an arbitrary $ " volume " \omega $ in the
relevant region of the phase space and let the $"surface"$ enclosing
this volume be denoted by $ \sigma. $ Then the rate at which the
number of  representative points in this volume increases with time
is written as:
\begin{eqnarray}
\frac{\partial}{\partial t} \int_{\omega}\rho d\omega
\end{eqnarray}
Where $ d\omega \equiv ( d^{3N}q d^{3N}p ). $On the other hand, the
$ net $ rate at which the representative points $ "flow" $ out of $
\omega $ (across the bounding surface $\sigma$ ) is given by:
\begin{eqnarray}
\int_{\sigma} \rho (\upsilon \cdot \hat{n})d
\sigma
\end{eqnarray}
Here,$\upsilon$ is the velocity vector of the representative points
in the region of the (outward) unit vector normal to this element.
By the divergence theorem(10) can be written as:
\begin{eqnarray}
\int_{\omega} div(\rho \upsilon)d\omega
\end{eqnarray}
of course the operation of divergence here means the following:
\begin{eqnarray}
div(\rho \upsilon) \equiv \sum^{3N}_{i = 1}
\{\frac{\partial}{\partial q_{i}}(\rho \dot{q}_{i}) +
\frac{\partial}{\partial p_{i}}(\rho \dot{p}_{i})\}
\end{eqnarray}
In view of the fact that are no $ "surfaces" $ or$ "sinks" $ in the
phase space and hence the total number of representative points
remains conserved\footnote{This means that in the ensemble under
consideration neither have any new members being admitted nor have
any old ones being expelled.}, we have, by (9) and (11),
\begin{eqnarray}
\frac{\partial}{\partial t} \int_{\omega} \rho d\omega = -
\int_{\omega} div(\rho \upsilon) d\omega
\end{eqnarray}
That is:
\begin{eqnarray}
\int_{\omega} \{\frac{\partial \rho}{\partial t} + div(\rho
\upsilon)\} d\omega = 0
\end{eqnarray}
Now, the necessary and sufficient condition that integral (14)
vanishes for all arbitrary volumes $\omega$ is that the integrand
itself vanishes $everywhere$ in the relevant region of the phase
space.Thus we must have:
\begin{eqnarray}
\frac{\partial \rho}{\partial t} + div(\rho \upsilon) = 0
\end{eqnarray}
Which is the$ equation of continuity$ for the swarm of the
representative points.Combining (12)and(15),we obtain:
\begin{eqnarray}
\frac{\partial \rho}{\partial t} + \sum^{3N}_{i=1} (\frac{\partial
\rho}{\partial q_{i}} \dot{q}_{i} + \frac{\partial \rho}{\partial
p_{i}}\dot{p}_{i}) + \rho  \sum^{3N}_{i=1}(\frac{\partial
\dot{q}_{i}}{\partial q_{i}} + \frac{\partial \dot{p}_{i}}{\partial
p_{i}}) = 0
\end{eqnarray}
The last group of terms vanishes identically because, by the
equation of motion, we have for all $i$,
\begin{eqnarray}
\frac{\partial \dot{q}_{i}}{\partial q_{i}} = \frac{\partial^{2}
H(q_{i},p_{i})}{\partial q_{i} \partial p_{i}} \equiv
\frac{\partial^{2} H(q_{i},p_{i})}{\partial p_{i}\partial q_{i}} = -
\frac{\partial \dot{p}_{i}}{\partial p_{i}}
\end{eqnarray}
Further,since $\rho \equiv \rho(q_{i},p_{i},t)$,the remaining terms
in (16) may be combined to form the $ total$ time derivative of
$\rho$,with the result that:
\begin{eqnarray}
\frac{d\rho}{dt} = \frac{\partial \rho}{\partial t} + [\rho , H] = 0
\end{eqnarray}
Equation (18)\footnote{ We recall that the $ poisson bracket$ $[\rho
, H]$ stands for the sum:
\\
$\sum^{3N}_{i=1} (\frac{\partial \rho}{\partial q_{i}}
\frac{\partial H}{\partial p_{i}} - \frac{\partial \rho}{\partial
p_{i}} \frac{\partial H}{\partial q_{i}})$} embodies the so called $
Liouvilles theorem$ (1838).According to this theorem, the $"local"$
density of the representative points,as viewed by an observer moving
with a representative points,stays constant in time.
\\
Now we can understand :
\begin{eqnarray}
L = i [H,]
\end{eqnarray}
where i[H,] is a Liouville operator , symbolled by $L$ and defined
as:
\begin{eqnarray}
iL = \sum^N_{i=1} (\frac{\partial H}{\partial p_i}
\frac{\partial}{\partial q_i} - \frac{\partial H}{\partial q_i}
\frac{\partial}{\partial p_i})
\end{eqnarray}
so we write:
\begin{eqnarray}
\nonumber \frac{\partial \rho}{\partial t} = [H,\rho]\\
\frac{\partial \rho}{\partial t} = - iL\rho
\end{eqnarray}
this has formal solution,
\begin{eqnarray}
\rho(t) =\exp(- iLt)\rho(0)
\end{eqnarray}
the classical Liouville equation is a highly non-trivial equation ,
where the momenta and coordinates of all the N-particles of system
are in principle coupled with each other.
\section{ii. Physical interpretation }
The quantity $ \rho(q^{(N)},p^{(N)},t) dp^{(N)}dq^{(N)}$ is the
probability that at a time  $t$  the physical system is in a
microscopic state corresponded by a phase point lying in the trivial
$6N$-dimensional phase space element $ dp^(N) dq^(N)$. Hence the
total number of systems in the ensemble is assumed by the integral
over phase space of distribution , $ \int \rho(p^{(N)},q^{(N)})
dp^{(N)} dq^{(N)}$.

The Liouville equation is the $6N$-dimensional analogue of equation
of persistence of an incompressible fluid.

It explains the fact that phase point of ensemble are neither
created nor destroyed. Thus we can interpret it as a conservation
theorem similar to charge conservation role in Electrodynamics.

\section{iii. Liouville theorems  }
The initial value problem for the Liouville equation was first
studied by Petrina and Gerasimenko  \cite{3} and then by Jiang
\cite{4},with its initial data in $L^1$ and $L^2$,respectively.
\\
It has been shown by using the operator semigroup theory that this
problem has a unique solution  if its  initial data belong to some
subset of $L^1$ or $L^2$ .
\subsection{A:CAUCHY PROBLEM FOR THE LIOUVILLE EQUATION}
The Liouville equation is an evolution:
\begin{eqnarray}
\frac{d}{dt} f_{N}(t) = - H^{N}_{\Lambda} f_{N} (t)
\end{eqnarray}
Where $ H^{N}_{\Lambda} $ is the infinitesimal operator of the group
$S^{N}_{\Lambda}(- t).$ Let us consider the liouville equation(23)
as an abstract evolution equation in the Banach space $
L_{N}(\Lambda)$ with initial conditions
\begin{eqnarray}
\nonumber f_{N}(0) = f^{0}_{N}.
\end{eqnarray}
THEOREM 1.A. The Cauchy problem for the Liouville equation (23) has
a unique solution in the space $ L_{N}(\Lambda) $ of summable
functions. It is given by the formula
\begin{eqnarray}
f_{N}(t) = S^{N}_{\Lambda}(- t) f^{0}_{N}.
\end{eqnarray}
For initial conditions $ f^{0}_{N} \in L^{0}_{N}(\Lambda) $ this
solution is a strong one and for arbitrary $ f^{0}_{N} \in
L_{N}(\Lambda) $ it is a generalized solution.
\\
PROOF.According to the well known result of functional analysis
\cite{103} the Cauchy problem for equation
\begin{eqnarray}
\nonumber (S^{N}_{\Lambda}(t)f_{N})(x_{1},...,x_{N}) =
\\
\nonumber f_{N}(X_{1}(t,x),...,X_{N}(t,x)) ,   (x_{1},...,x_{N})\in
\Gamma^{N}\backslash W^{0}_{N}
\\
\nonumber 0             ,            (q_{1},...,q_{N}) \in W_{N}.
\end{eqnarray}
has the
unique solution in $ L_{N}(\Lambda)$ which is given by (24).We shall
demonstrate that in the formula (24) defines a generalized solution.
For this purpose we consider the functional
\begin{eqnarray}
\nonumber(\varphi,f_{N}(t)) = \int dx_{1}...dx_{N} \varphi(x_{1},...,x_{N})f_{N}(t,x_{1},...,x_{N})
\end{eqnarray}
Where $\varphi$ is a continuously differentiable function with a
compact support that vanishes like functions from
$L^{0}_{N}(\Lambda)$ in some neighborhood of forbidden
configurations and $\partial \Lambda.$ Since $\varphi$ is bounded
and $ f_{N}(t) $ is summable ,the functional $ (\varphi,f_{N}(t)) $
exists.Using (24) and the fact that $ S^{N}_{\Lambda}(-t)$ is
isometric we transform $(\varphi,f_{N}(t))$ to the form
\begin{eqnarray}
\nonumber(\varphi,f_{N}(t)) = (S^{N}_{\Lambda}(t)\varphi,f^{0}_{N})
= \int dx_{1}...dx_{N}
(S^{N}_{\Lambda}(t) \varphi)(x_{1},...,x_{N})f^{0}_{N}(x_{1},...,x_{N}).
\end{eqnarray}
As in Theorem 1.A  we can show that the function $
(S^{N}_{\Lambda}(t)\varphi)(x_{1},...,x_{N})$ is differentiable with
respect to $t$, and expression
\begin{eqnarray}
\nonumber|\frac{1}{\Delta t}(S^{N}_{\Lambda}(t + \Delta t)\varphi -
S^{N}_{\Lambda}(t)\varphi) - \sum^{N}_{i=1} P_{i} \cdot
\frac{\partial}{\partial q_{i}}
S^{N}_{\Lambda}(t)\varphi|\longrightarrow 0
\end{eqnarray}
as $\Delta t\longrightarrow 0$ uniformly with respect to $
(x_{1},...,x_{N})$ belonging to compact sets. The functional
$(\sum^{N}_{i=1}P_{i} \cdot \frac{\partial}{\partial
q_{i}}(S^{N}_{\Lambda}(t)\varphi),f^{0}_{N})$, since the function
$\sum^{N}_{i=1}P_{i} \cdot\frac{\partial}{\partial
q_{i}}(S^{N}_{\Lambda}(t)\varphi)(x) $ is bounded. Therefore we can
differentiate with respect to $ t $ under the integral sign in the
functional $(S^{N}_{\Lambda}(t)\varphi,f^{0}_{N})$.As a result we get
\begin{eqnarray}
\nonumber \frac{d}{dt}(\varphi,f_{N}(t)) = (\sum^{N}_{i=1}P_{i}\cdot
\frac{\partial}{\partial
q_{i}}(S^{N}_{\Lambda}(t)\varphi),f^{0}_{N}) =
(S^{N}_{\Lambda}(t)\sum^{N}_{i=1}P_{i}\cdot \frac{\partial}{\partial
q_{i}}\varphi,f^{0}_{N}) = (\sum^{N}_{i=1}P_{i}\cdot \frac{\partial}{\partial q_{i}} \varphi,f_{N}(t))
\end{eqnarray}
and these equalities mean that the function $f_{N}(t) =
S^{N}_{\Lambda}(-t)f^{0}_{N}$ is a generalized (weak) solution of the Cauchy problem
for the Liouville equation (23).
\\
Theorem (1.A) holds also for a system of particles in $\Re^{v}.$ We
only have to replace $ S^{N}_{\Lambda}(-t)$ and $ H^{N}_{\Lambda}$
by their counterparts $ S^{N}(-t)$ and $ H^{N} $ in $L_{N}.$
\\
Let us construct the Liouville equation starting from the
Hamiltonian dynamics of a system of elastic balls. As well known
\cite{104} the class of pure states of Hamiltonian system of
particles can be described by the distribution function
\begin{eqnarray}
D_{N,\Lambda}(t,x_{1},...,x_{N}) = \sum_{\{i_{1},...,i_{N}\}}
\prod^{N}_{k=1} \delta(x_{ik} - X_{k}(t,x^{0}))
\end{eqnarray}
Where $X(t,x^{0})$ is the phase point that is occupied by a system
of $N$ particle at time moment $t$ if at the initial time moment $t
= 0$ the system was at phase point $ x^{0} \in \Re^{vN} \times
(\Lambda^{N}\backslash W_{N}).$ Here the summation with respect to
$\{i_{1},...,i_{N}\}$ is carried out over all the permutations of
the indices $ i_{k}\in (1,...,N).$ All possible states of the system
considered can be described by the distribution function on the
phase space $\Gamma^{N}(\Lambda) $:
\begin{eqnarray}
D_{N,\Lambda}(t,x_{1},...,x_{N}) = \int_{\Gamma^{N}(\Lambda)}
dx^{0}_{1}...dx^{0}_{N}D^{0}_{N}(x^{0}_{1},...,x^{0}_{N})\sum_{\{i_{1},...,i_{N}\}}
\prod^{N}_{k=1} \delta (x_{ik} - X_{k}(t,x^{0}))
\end{eqnarray}
Which is superposition of the pure states (25) of the system with a
definite weight $D^{0}_{N}(x^{0}_{1},...,x^{0}_{N}).$ The last
quantity is a known function of the distribution of particle
coordinates and momenta that is specified at the initial time moment
$ t = 0 .$ Consequently an initial state of a system of $N$ elastic
balls is , in general, described by the function
\begin{eqnarray}
D_{N,\Lambda}(0,x_{1},...,x_{N}) =
\int_{\Gamma^{N}(\Lambda)}dx^{0}_{1}...dx^{0}_{N}D^{0}_{N}(x^{0}_{1},...,x^{0}_{N})\sum_{\{i_{1},...,i_{N}\}}
\prod^{N}_{k=1} \delta (x_{ik} - x^{0}_{k})
\end{eqnarray}
Which is equal to
\begin{eqnarray}
\nonumber D_{N,\Lambda}(0,x_{1},...,x_{N}) = N!
D^{0}_{N}(x_{1},...,x_{N}).
\end{eqnarray}
We note that the distribution function $ D_{N,\Lambda}(t)$ has the
following properties:
\begin{eqnarray}
D_{N,\Lambda}(t,x_{1},...,x_{N}) = D_{N,\Lambda}
(t,x_{1},...;q_{i},p^{\ast}_{i};...;q_{j},p^{\ast}_{j};...,x_{N}),
\\
\nonumber |q_{i} - q_{j}| = d, i\neq j \in (1,...,N),
\\
\nonumber D_{N,\Lambda}(t,x_{1},...,x_{N}) = D_{N,\Lambda}
(t,x_{1},...;q_{i},p^{\ast}_{i};...,x_{N}),q_{i} \in
\partial \Lambda,
\\
\nonumber D_{N,\Lambda}(t,x_{1},...,x_{N}) = 0 , (q_{i},...,q_{N})
\in W_{N}.
\end{eqnarray}
If we change the variables $x^{0} \longrightarrow X(-t,x)$ in the
integrand of (26),then after calculating the integrals we get the
relation
\begin{eqnarray}
D_{N,\Lambda}(t,x_{1},...,x_{N}) =
D_{N,\Lambda}(0,X_{1}(-t,x),...,X_{N}(-t,x)).
\end{eqnarray}
From this formula and definition (2.3) of the evolution operator is
follows that the state at an arbitrary time moment is connected with
the initial state by the following formula:
\begin{eqnarray}
D_{N,\Lambda}(t,x_{1},...,x_{N}) =
(S^{N}_{\Lambda}(-t)D_{N,\Lambda}(0))(x_{1},...,x_{N}).
\end{eqnarray}
In order to construct an equation that determines the evolution of a
state of a system of elastic balls we differentiate the right hand
side of (30) with respect to time for $ D_{N,\Lambda}(0) \in
L^{0}_{N}(\Lambda).$
\\
From Theorem 1.A it follows that
\begin{eqnarray}
\frac{\partial}{\partial t} D_{N,\Lambda}(t,x_{1},...,x_{N}) = -
\sum^{N}_{i=1} P_{i}\cdot \frac{\partial}{\partial q_{i}}
D_{N,\Lambda}(t,x_{1},...,x_{N})
\end{eqnarray}
With the corresponding boundary condition in $Poisson bracket$ and
condition (28 ). The relation (31) is obtained as consequence of
(29). From now on we shall treat (31) as an equation for an unknown
function $ D_{N,\Lambda}(t,x_{1},...,x_{N}),$ which is the Liouville
equation. We have to add to it initial and boundary condition (28),
which are satisfied by$ D_{N,\Lambda}(t,x_{1},...,x_{N}).$
\subsection{B:ON THE LIOUVILLE THEOREM}
Headword to study our problem by using  the operator semigroup
theory,equation (8) can be rewritten as
\begin{eqnarray}
\nonumber \frac{d \rho}{d t} + A \rho = 0
\end{eqnarray}
Where $A \triangleq - [H,.]$. $A$ can be in regarded to as either a
distributional differential operator in $L^{2}({\bf R})^{2k}$ or a
mapping of $L^{2}({\bf R})^{2k}$ into itself.
\\
THEOREM 1.B .operator $
A $ is the infinitesimal generator of a contractive operator
semigroup.
\\
\\
In order to prove Theorem 1.B , we'll first show the following
lemmas.
\\
\\
LEMMA 1.B . $ A $ is an increasingly-generated operator , i.e. , $
Re(Au,u)\geq 0 $ for $ \forall u \in D(A)$ ,where $ (.,.) $ denotes
the internal product in $ L^2( R^{2N} ). $
\\
PROOF.By equation(8),we know that the for $\forall u, \nu \in D(A),$
\begin{eqnarray}
\nonumber (A\nu,u) = - \int \overline{u} \sum^{N}_{i=1}
(\frac{\partial H}{\partial q_{i}} \frac{\partial \nu}{\partial
p_{i}} - \frac{\partial \nu}{\partial q_{i}} \frac{\partial
H}{\partial p_{i}}) dq dp.
\end{eqnarray}
By integrating partially and using  the Hamiltonian equation ,the
above integral can also be written as:
\begin{eqnarray}
\nonumber (A \nu,u) = \int \nu \sum^{N}_{i=1} (\frac{\partial
H}{\partial q_{i}} \frac{\partial \overline{u}}{\partial p_{i}} -
\frac{\partial \overline{u}}{\partial q_{i}} \frac{\partial
H}{\partial p_{i}}) dq dp
\\
\nonumber = \overline{\int \overline{\nu} \sum^{N}_{i=1}
(\frac{\partial H}{\partial q_{i}} \frac{\partial u}{\partial p_{i}}
-\frac{\partial u}{\partial q_{i}} \frac{\partial H}{\partial
p_{i}}) dq dp } = -\overline{(Au,\nu)}
\end{eqnarray}
In particularly,
\begin{eqnarray}
\nonumber (Au,u) = -\overline{(Au,u)}
\end{eqnarray}
Or equivalently, $Re (Au,u) = 0, $ for $ \forall u \in D(u).$
\\
\\
LEMMA 2.B . $ A $ is a closed operator with dense domain in $ L^2 (
R^{2N}). $
\\
\\
LEMMA 3.B. $ \lambda I + A $ is mapping of $ L^2 (R^{2N}) $ on to it
self , i.e., the rang $ R( \lambda I + A ) $ of the mapping $
\lambda I + A $ is $ L^2 ( R^{2N}) $ , for any given $ \lambda > 0 $
, where $ I $ is identity mapping.
\\
\\
For $ \forall \chi \in D(A). $ Then $ -A $ is the infinitesimal
generator of $ S(t). $
\\
Similarly, we can also show that
\\
THEOREM 2.B. operator $ -A $ is the infinitesimal generator of a
contractive operator semigroup.
\\
\\
According to the operator semigroup theory , by Theorem 1.B , we
define
\\
THEOREM 3.B. If $ u_{0} \in D(A) $ then the initial value problem $
[ ( \frac{du}{dt} + Au = 0) , ( u_{|t=0} = u_{0} )] $ has a unique
solution $ u = S(t) u_{0} \in C^{1} ( [0, +\infty);L^2(R^{2N})) \cap
C^{0} ([0 , +\infty);D(A)) $ where $ S(t) $ is defined by Eqs. $ [
(S(t) = \lim_{ \lambda \longrightarrow +\infty} S_{\lambda} (t)) ; (
S_{\lambda} (t) = \exp^{A_{k}t} ) ; ( A_{\lambda} \chi =
[\lambda^{2} (\lambda I + A )^(-1) - \lambda I]\chi) ; for \forall
\chi \in L^2(R^{2N}) and , as \lambda \rightarrow +\infty
;(A_{\lambda} \chi \rightarrow -A\chi)]$
\\
\\
THEOREM 4.B. If $ u_{0} \in D(A^{k})$ , $ k $ is a positive integer
, then the solution $ u $ of Eqs. $ [( \frac{du}{dt} + Au = 0) , (
u_{|t=0} = u_{0} )] $ belongs to $ \cap^{k}_{j = 0} C^{k-j}
([0,+\infty);D(A^{j}) where , D(A^{0})\triangleq L^2(R^{2N}).$
\\
For study their proofs refer \cite{4}.
\\
\\
The total time derivative of the phase space density  is denoted
by$\frac{d \rho}{d t} $. Then:
\begin{eqnarray}
\frac{d \rho}{dt} = \frac{\partial \rho}{\partial t} +
\sum^{N}_{i=1} [ \frac{\partial \rho}{\partial q_{i}} \frac{\partial
q}{\partial t} + \frac{\partial \rho}{\partial p_{i}} \frac{\partial
p}{\partial t}] = 0
\end{eqnarray}
Where $ \frac{d \rho}{dt} = 0 $ , thus we provide the Liouville
Theorem  in classical Liouville equation  : In conservative system
the distribution function is constant along any trajectory in phase
space.
\\
\\
The Petrina and Gerasimenko survey contains more and sound physical
description in related to the operational approach to the solution
of Liouville equation and non-linear ones(Bogolyubov). The Jiang
argument constructed  a weak group  based on the non-differential
operator version of the Liouville equation and the result as  we can
guess from the elementary group theory is a semi group represents a
formal solution.AS  Petrina and Gerasimenko discussed, this formal
solution can be rewritten as a familiar Dyson series in terms of the
normal generators of a complete group.However,since $L^1({\bf
R}^{2K})$ and $L^2({\bf R}^{2k})$ have not any mutual inclusion
relation,the two existence results are different and the $L^2$
existence theorem is not a theoretical scrutiny of the $L^1$ one.
\subsection{ C:SOLUTION OF BOGOLYUBOV EQUATION}
The Bogolyubov equations for a system of elastic balls are the
evolution equations
\begin{eqnarray}
\frac{d}{dt}F(t) = - H F(t) + d^{2}A F(t)
\end{eqnarray}
Where $-H + d^{2}A$ is the infinitesimal operator of the group
$U(t).$ Let us consider the Bogolyubov equation (34) as abstract
evolution equations in the Banach space $L$ with the initial
conditions
\begin{eqnarray}
\nonumber F(0) = F^{0}
\end{eqnarray}
It can be show that \cite{3} \cite{104},\cite{103} that Cauchy
problem (34) for the Bogolyubov equations has in the space $L$ of
sequence of summable functions a unique solution that is given by
the formula
\begin{eqnarray}
F(t) = U(t) F^{0} =\exp(\int dx)S(-t)\exp(\int dx) F^{0}.
\end{eqnarray}
In components (35) can be written as follows:
\begin{eqnarray}
F_{s}(t,x_{1},...,x_{s}) = \sum ^{\infty}_{n=0} \sum^{n}_{k=0}
\frac{(-1)^{k}}{k!(n - k)!} \int
dx_{s+1}...dx_{s+n}S^{s+n-k}(-t,x_{1},...,x_{s+n-k})\times
F^{0}_{s+n}(x_{1},...,x_{s+n}).
\end{eqnarray}
Each term on the right hand side of (36) is well defined, since the
integrand is defined almost everywhere outside $M^{0}_{s+n}$, and
the series converges in the metric of space $L_{s}$ for $t \in
]-\infty,+\infty[.$
\\
THEOREM 1.C. The Cauchy problem for the Bogolybov equations (34) has
the unique solution in $L$ that is given by (35).For initial
conditions $ F^{0} \in L^{0}$ this solution is a strong solution and
for arbitrary $ F^{0} \in L$ it is a generalized (weak) solution.
\\
For study its proof refer\cite{3}.
\\
We note that the evolution equation
\begin{eqnarray}
\nonumber \frac{d}{dt} \varphi(t) = H \varphi(t) +
d^{2}A^{\ast}\varphi(t)
\end{eqnarray}
Are called the dual Bogolyubov equations \cite{104},\cite{105}.
\\
Obviously Theorem 1.C holds also for the Bogolyubov equations that
describe a system of particles moving in $\Lambda$:
\begin{eqnarray}
\frac{d}{dt}F_{\Lambda}(t) = - H_{\Lambda}F_{\Lambda}(t) + d^{2} A
F_{\Lambda}(t)
\end{eqnarray}
Where $-H + d^{2}A$ is the infinitesimal operator of the group
$U_{\Lambda}(t)$ in$(U(t) = \exp(\int dx)S(-t)\exp(-\int dx))$. The
solution of Cauchy problem for equations (37) is defined by
\begin{eqnarray}
F_{\Lambda}(t) = U_{\Lambda}(t)F^{0}_{\Lambda} = \exp(\int_{\Lambda}
dx) S_{\Lambda}(-t) \exp(-\int_{\Lambda} dx) F^{0}_{\Lambda}.
\end{eqnarray}
We not that the rigorous derivation of the Bogolyubov equations
(34)((37)) for a system of elastic balls,as an evolution equation in
the space $L(L(\Lambda))$ of sequences of summable functions with
the infinitesimal operator $ -H + d^{2} A$ of the group $U(t).$
Another method of justifying Bogolyubov equations consists in
constructing these equations starting from Liouville equations (23).
\section{iv. The stochastic Liouville equation  }
The transport equation is the stochastic Liouville equation ( SLE
).[5-9]
\\
It was shown,\cite{10}, that the SLE can be re-expressed as
presenting a \emph{formal}trapping problem in a manner analogous to
that employed in the theory of mutual annihilation of exciton's
developed by one of the present authors.\cite{11}
\\
The exact solution is particulary appropriate as a starting point
for the study of the scattering function relevant to experiments
involving probe particles such as neutrons.The particular feature
that their solution,\cite{10},possesses is the ability to address
the degree of transport \emph{coherence} of the moving particle
which produces the scattering line shape.
\\
Before we explain section V we want to deal with spins which are
in random motion.
\subsection{A:THE DENSITY MATRIX $p(\Omega)$}
The simplest example of such a type of motion is that of spins which
jump back and forth among a number of sites with different chemical
or  magnetic environment.They \cite{12},label the sites with the
numbers $ \upsilon = 1,2,...,n $ and define the distribution vector
$ p_{\upsilon}(t) $ as the vector of the probabilities of finding
spins in the sites $ \upsilon $ at the time $ t $.Another example is
the case of translational motion.
\\
The position of a spin is then characterized by the spatial
coordinate $ \vec{r} $ and the distribution function is $
p(\vec{r},t) .$ The case of rotatory motion of molecules is very
common. The orientation of the molecules is then defined by the
Eulerian angles,$ \Omega .$All these cases in one formalism and
choose $ \Omega $ as notation of the random coordinate.Thus the
function $ p(\Omega , t)$ will denote the probability density of
finding spins in the environment characterized by a particular $
\Omega .$
\\
The spins can satisfactorily be described in terms of elementary
magnets $ \vec{\vec{m}}, $which can differ in direction but not in
magnitude $ |m|. $
\\
The state of the whole spin system is then given by the combined
probability density $ P(\Omega , \vec{m})$,which we take to be
normalized:
\begin{eqnarray}
\int\int P(\Omega ,\vec{m})d\Omega d\vec{m} = 1
\end{eqnarray}
From this function we can extract the following quantities;the
molecular distribution function:
\begin{eqnarray}
p(\Omega) = \int P(\Omega ,\vec{m})d\vec{m}
\end{eqnarray}
The distribution function of the magnetization:
\begin{eqnarray}
f(\vec{m}) = \int P(\Omega ,\vec{m})d\Omega
\end{eqnarray}
The average magnetic moment:
\begin{eqnarray}
\langle\vec{m}\rangle_{0} = \int\int \vec{m} P(\Omega
,\vec{m})d\Omega d\vec{m}
\end{eqnarray}
The macroscopic magnetization of the sample:
\begin{eqnarray}
\vec{M}_{0} = N\langle\vec{m}\rangle_{0}
\end{eqnarray}
Where $N$ is the total number of spins per unit volume; and
finally the quantities:
\begin{eqnarray}
\langle\vec{m}(\Omega)\rangle = \int \vec{m}P(\Omega,\vec{m})d\vec{m}
\end{eqnarray}
and
\begin{eqnarray}
\vec{M}(\Omega) = N\langle\vec{m}(\Omega)\rangle
\end{eqnarray}
Which may be called the $\Omega-dependent$ magnetization density.
\\
These last two quantities have the following physical
meaning.Suppose that we could measure separately the total
magnetization of the molecules which have their orientations between
$\Omega$ and $ \Omega + d\Omega .$ Their measurement\cite{12}, would
then yield $ \vec{M}(\Omega)d\Omega .$or suppose that we could
measure the total magnetization of the sample, but with an
$\Omega-dependent$ weight $ w(\Omega) .$We would then measure $ \int
w(\Omega)\vec{M}(\Omega)d\Omega . $We can define $\langle
\vec{m}(\Omega)\rangle $ in a slightly different way by writing:
\begin{eqnarray}
P(\Omega ,\vec{m}) = p(\Omega)q(\Omega ,\vec{m})
\end{eqnarray}
Where $ q(\Omega ,\vec{m}) $is the normalized probability density of
finding the magnetic moment at the value $\vec{m}$, provided that we
know that the molecule is in $\Omega .$
We then have:
\begin{eqnarray}
\langle \vec{m}(\Omega)\rangle = p(\Omega)\int
\vec{m}q(\Omega,\vec{m})d\vec{m}
\end{eqnarray}
This notation emphasizes that $\langle \vec{m}(\Omega)\rangle $ is
the average magnetic moment of the molecules with orientation
$\Omega$,multiplied by the probability of finding the molecules in $
\Omega .$Thus $ \langle \vec{m}(\Omega)\rangle $contains information
on the internal magnetic state of the molecules and on the molecular
distribution in $ \Omega-space .$It is important to appreciate this
point when the theory is applied to systems where $ p(\Omega)$ is not
a uniform function, as is the case with chemical exchange between unequally
populated sites, or with partially oriented molecules.
\\
It is not necessary to say, in the case of jumps the parameter$
\Omega $is replaced by an index $\nu$ and the integrations over
$\Omega$ became summations.
\\
We want consider molecular spin systems which are not adequately
represented by a single magnetic moment $\vec{m}.$ The state of the
spins has then to be described quantum mechanically by the wave
function $\psi$ or the density matrix $\rho.$As for the density
matrix,we should distinguish between $\rho$ of each separate spin
system whose matrix elements $\rho_{nm}$ are the products $
a_{n}a^{\ast}_{m}$ of the coefficient in the expansion of $\psi$ and
its ensemble average,which we denote by $ \rho^{-1,2} .$If the spin
system consists of a single spin $I = \frac{1}{2}$, there is a one
to one correspondence between $\vec{m}$ and $\rho$, and between
$\langle \vec{m}\rangle$ and $\rho^{-3}.$However,if the spin system
has more than two levels,this comparison can not always be
made.Nevertheless it is often convenient to visualize the behavior
of complicated spin systems through a model of magnetization. They
utilized\cite{12} this analogy and assume a combined probability
density $P(\rho,\Omega)$\cite{13},from which we drive the
distribution function of the density matrices:
\begin{eqnarray}
f(\rho) = \int P(\Omega,\rho)d\Omega
\end{eqnarray}
The function $f$ and $P$ are defined in the space of all the
possible density matrices $\rho.$This space is restricted to $\rho
s$ with matrix elements which are products of $a_{n}$ and $
a^{\ast}_{m}$ taken from normalized sets of coefficients $a_{n}.$
These are the matrices which satisfy the conditions of so called
pure states. Not that the space of $\vec{m}$ was also restricted to
vectors with a constant modulus $|m|.$
\\
The determination of the functions $ P(\Omega,\rho) $ and $ f(\rho) $
is very difficult.
\\
We come to the main subject of introduction. Somewhat inexactly,call
the $ \Omega-dependent $spin density matrix
\begin{eqnarray}
\bar{\rho}(\Omega) = \int \rho P(\Omega,\rho)d\rho
\end{eqnarray}
Which is the analogy of $\langle \vec{m}(\Omega)\rangle .$It may be
used to compute the ensemble average of the expectation value of any
physical quantity $Q(\Omega).$Which also depends on $\Omega$,
\begin{eqnarray}
\nonumber\langle Q\rangle = \int tr[\rho
Q(\Omega)]P(\Omega,\rho)d\Omega d\rho
\\
=\int tr[\bar{\rho}(\Omega)Q(\Omega)]d\Omega
\end{eqnarray}
In most experiment performed on the spin system we measure a
property $ Q $ which is independent of $\Omega.$ It is then
sufficient to know $\bar{\rho}_{0}$ in order to calculate:
\begin{eqnarray}
\langle Q\rangle = tr Q\bar{\rho}_{0};\bar{\rho}_{0} = \int
\bar{\rho}(\Omega)d\rho
\end{eqnarray}
By analogy with equation(47) we can also write:
\begin{eqnarray}
\bar{\rho}(\Omega) = p(\Omega)\int \rho q(\Omega,\rho)d\rho =
p(\Omega)\bar{\rho}_{\Omega}
\end{eqnarray}
Which emphasizes that $\bar{\rho}(\Omega)$ is the product of local
ensemble average $\bar{\rho}_{\Omega}$ and the probability of
finding molecules in $\Omega.$
\subsection{B:THE DENSITY MATRIX$\rho(t)$ AND SLE}
Let us now assume that the interaction of the spins are describe by
an $\Omega-dependent$ Hamiltonian $H(\Omega)$,i.e. a Hamiltonian
which is different for spins belonging to molecules with different
points.How,then,does $P(\Omega,\rho,t)$ develop in time if the
molecules are moving randomly in $\Omega-space$?
\\
One starts to look at one particular spin system in the ensemble. Since this system belongs to
a molecule which moves rapidly from one $\Omega$ to the other, a
time-dependent Hamiltonian $H(t)$ is observed by the spins.$H(t)$ is
now written as the sum of a constant $H_{0}$ and a time-dependent
local Hamiltonian $H'(t)$ with vanishing time average.
\\
The density matrix of spin system in question changes according to this
Hamiltonian:
\begin{eqnarray}
(\frac{d}{dt})\rho(t) = \frac{i}{\hslash}[\rho(t),H_{0}] +
\frac{i}{\hslash}[\rho(t),H'(t)]
\end{eqnarray}
This equation is first solved by following a perturbation
treatment,and then the ensemble average is taken.This yields the
master equation. An essential point in the averaging procedure is
that the correlation $\rho(t)$ and $H(t)$ is neglected.This is only
permissible if the correlation times $\tau$ of the matrix elements
of $H'(t)$ are so short that:
\begin{eqnarray}
|H'(t)|\tau \ll \hslash
\end{eqnarray}
It is important to stress the difference between the density
matrices $\rho(t)$ and $\bar{\rho}(\Omega). \rho(t)$ is a pure
state\cite{12} and $\bar{\rho}(\Omega)$ is an ensemble average.
Equation(53) is a stochastic equation which defines the stochastic
process $\rho(t)$ in terms of the stochastic process $ H(t).$ More
precisely by writing:
\begin{eqnarray}
(\frac{d}{dt})\rho(t) = \frac{i}{\hslash}[\rho(t),H_{0} +
H'\{\Omega(t)\}]
\end{eqnarray}
We see that process $\rho(t)$ actually depends on the stochastic
process $\Omega(t).$ In most problems dealt whit in magnetic
resonance theory some model is assumed for the description of
stochastic process $\Omega(t).$In nearly all cases this is a
stationary $Markoffian$ process.It is then assumed that the
probability density $p(\Omega,t)$ satisfies the equation:
\begin{eqnarray}
(\frac{\partial}{\partial t})p(\Omega,t) = \Gamma p(\Omega,t)
\end{eqnarray}
Where $\Gamma$ is a time-independent $Markoffian$ operator,
operating on functions of $\Omega.$More generally,$\Omega(t)$ is the
projection of a $Markoffian$ process,i.e. $\Omega$ should be
supplemented whit additional variables to form a complete set of
random variables which make a $Markoffian$ process. In order to
retain a simple notation we assume that $\Omega$ it self is a
$Markoffian$ process.Equation(56) fits well in formalism of
section(IV.A) but it can be less directly applied to further
development (53). Thus we follow $Kubo's$ development\cite{13} and
write formally for the rate equation of $P(\Omega,\rho,t)$
\begin{eqnarray}
\frac{\partial}{\partial t}P(\Omega,\rho,t) = \{
-\frac{\partial}{\partial \rho}\frac{i}{\hslash}[\rho,H(\Omega)] +
\Gamma \}P(\Omega,\rho,t)
\end{eqnarray}
This can be regarded as a composite $Markoffian$ process. It is a
coarse-grained description of complete Liouville equation of the
density of states of combined system of lattice and spins,utilizing
the stochastic property $\Gamma.$ Thus we may call it a stochastic
Liouville equation\cite{13,14,15,16,17,18}.
\\
We first multiply equation (57) by $\rho$ and integrate over $\rho.$
This yields,with equation (49),
\begin{eqnarray}
(\frac{\partial}{\partial t})\bar{\rho}(\Omega,t) =
\frac{i}{\hslash}[\bar{\rho}(\Omega,t),H(\Omega)] + \Gamma
\bar{\rho}(\Omega,t)
\end{eqnarray}
This is the equation which we shall refer to as the $SLE$.
\\
In this derivation of $SLE$ an important approximation has been
made.It is assumed that the molecules execute their random motions
regardless of the state in which the spins find themselves.Thus the
reaction of spin system to its surroundings is ignored.In other
words we neglected the energy exchange between the lattice and the
spins. However,as $Kubo stated$\cite{13},this is permissible for
instance when the temperature of the bath is sufficiently high
compared with the possible energy exchange. Many examples in NMR or
$Mossbauer$ effects belong to this category because the reaction to
the molecular motion of the bath is extremely small. Thus equation
(58) has a wide range of application in line shape problems in
magnetic resonance\cite{18}.
\\
In the case of jumps the operator $\Gamma$ is a matrix and the $SLE$ takes the
form:
\begin{eqnarray}
(\frac{d}{dt})\bar{\rho}_{\nu} =
\frac{i}{\hslash}[\bar{\rho}_{\nu},H_{\nu}] + \sum_{\mu}
\Gamma_{\nu\mu}\rho_{\mu}
\end{eqnarray}
Where $\Gamma_{\nu\mu}$ are reciprocals of mean residence times.
\section{v. SLE and method of solution  }
The stochastic Liouville equation in its simplest from is given by:
\begin{eqnarray}
\dot{\rho}_{mn} = -i[H,\rho]_{mn} -\alpha(1 - \delta_{mn})\rho_{mn}
\end{eqnarray}
and describes the time evolution of density matrix $ \rho $ of the
moving particle in the representation of site states $ m,n $. The
system in which the particle moves is a crystal, i.e.,possesses
translational periodicity. The intersite interaction is $ H $. The
last in $ (60) $ describes the randomizing process whereby the
off-diagonal elements of $ \rho $ decay, $ \alpha $ being the rate
at which this process occurs.Alternatively, $ \alpha $ may be looked
upon as the average rate of scattering among the band states of the
particle. In the limit of no scattering, $ (60) $ describes coherent
motion whereas, in the opposite limit it describes hopping or
incoherent motion.
\\
The indices $ m,n $ are vectors in the appropriate number of
dimensions. A different form of the SLE is:
\begin{eqnarray}
\dot{\rho}_{mn} = -i[H,\rho] -\alpha (1 -\delta_{mn})\rho_{mn} +
2\delta_{mn} \sum_{r} (\gamma_{mr} \rho_{rr} - \gamma_{rm}
\rho_{mm})
\end{eqnarray}
and describes in addition to the processes included in $ (60) $, a
transport channel wherein the particle hops from sites $ r $ to
sites $ m $ at rates $ 2\gamma_{mr} $.
\\
The $ SLE (60) $ can be looked upon as presenting a formal trapping
problem in a 2d-dimensional space where $ d $is the dimensionality
of the system under analysis. We should use $ \eta $ to denote the
\emph{homogeneous}solution of $ (60) $, i.e.,its solution in the
last term:
\begin{eqnarray}
\eta_{mn} (t) = \sum_{m'n'} (\psi_{m - m',n -n'}(t) \rho_{m'n'}(0)).
\end{eqnarray}
The quantities $ \psi $ are the density-matrix propagators, i.e.,
the solution of $ (60) $ for $ \alpha = 0 $ for the initial
conditions $ \rho_{mn} (0) = \delta_{m 0} \delta_{n 0}. $ If we
recast $ (60) $ as:
\begin{eqnarray}
\dot{\rho}_{mn} + \alpha \rho_{mn} = -i[H,\rho]_{mn} + \alpha
\delta_{mn} \rho_{mn}
\end{eqnarray}
We see that $ \alpha $ produces two perturbations on $ (62) $ : that
caused by $ (\alpha \rho_{mn}) $ on the left side and that caused by
$ \alpha \delta_{mn} \rho_{mm} $ on the right side.
\\
In the context of the dynamics of a hypothetical walker whose
unperturbed motion is given by $ (62) $ , these two terms represent,
respectively,an overall decay akin to the radiative decay of a
moving exciton,\cite{7}, and a trapping or annihilation
process,[7-9,11-19],which takes place only when $ m = n, $i.e, in a
special trap-influenced region in the $ m,n $ space.
\\
The first of the two terms introduces a simple multiplicative factor
in to the solutions $ (62) $ , but the second requires an analysis
through the defect technique,\cite{19}.
\\
Equation $ (63) $ takes the form:
\begin{eqnarray}
\tilde{\rho}_{mn} (\epsilon) = \tilde{\eta}_{mn} (\epsilon + \alpha)
+ \alpha \sum_{m'} (\tilde{\psi}_{m-m',n-m'} (\epsilon + \alpha )
\tilde{\rho}_{m'm'}(\epsilon))
\end{eqnarray}
in the Laplace domain where $ \epsilon $ is the Laplace variable and
tildes denote Laplace transforms.
\\
The case $ m = n $ gives:
\begin{eqnarray}
\tilde{\rho}_{mm} (\epsilon) = \tilde{\eta}_{mm} (\epsilon + \alpha)
+\alpha \sum_{m'} (\tilde{\psi}_{m -m',m-m'} (\epsilon + \alpha)
\tilde{\rho}_{m'm'} (\epsilon))
\end{eqnarray}
which involves only diagonal elements of the density matrix in the
representation of $ m,n. $ We solve $ (65) $ through the use of
discrete Fourier transforms.
\\
These are defined through relations such as:
\begin{eqnarray}
\rho^{k} = \sum_{m} \rho_{mm} \exp( ikm )
\end{eqnarray}
where $ k $ is generally a vector and $ km $ a dot produce.
\\
The result is:
\begin{eqnarray}
\tilde{\rho}^{k} (\epsilon) = \frac{\tilde{\eta}^{k} (\epsilon +
\alpha)}{1-\alpha \tilde{\psi}^{k} (\epsilon + \alpha )}.
\end{eqnarray}
It is straightforward to substitute $ (67) in (64). $
\\
One then finds that the solution of the $ SLE (60) $ is given by $
(62) $ with the replacement:
\begin{eqnarray}
\tilde{\psi}_{m-m',n-n'}
(\epsilon)\rightarrow\tilde{\psi}_{m-m',n-n'} (\epsilon) +
\frac{\alpha}{N} \sum_{r,s,k} [\frac{\exp(ik(s-r))}{1 - \alpha
\tilde{\psi}^{k} (\epsilon + \alpha)}] \tilde{\psi}_{m-r,n-r}
(\epsilon + \alpha)\tilde{\psi}_{s-m',s-n'} (\epsilon+\alpha).
\end{eqnarray}
This result is exact and explicit. It is explicit in the sense that
once one knows the $ \psi's $ ,i.e., the propagators of $ (60) $ or
$ (63) $ in the absent of $ \alpha $, one can write down the
solutions of $ (60) $ by following the prescription of $ (68) $ and
the right-hand side of $ (62) $ , for arbitrary initial conditions.
\\
The practical usefulness of $ (68) $ depends on the simplicity, or
lack thereof, of the quadrature problem involved in the inversions
of the transforms.
\\
\section{vi. Classical Liouville Theory and the Microscopic
Interpretation of Black Hole Entropy}
 It has been computed that the entropy of
a Schwarzschild black hole could be derived by finding a classical
central charge of the Virasoro algebra of a Liouville theory and
using the Cardy formula. This is done by performing a dimensional
reduction of the Einstein–Hilbert action with the Ansatz of
spherical symmetry and writing the metric in conformally flat
form.Near horizon approximation the field equation for the conformal
factor decouples becoming a Liouville equation.This computation is
independent from a specific quantum theory of gravity
model\cite{20}.In order to understand the role of 2-D conformal
symmetry in counting the black hole microstates let us recall some
standard features of 2-D conformal symmetry. By introducing complex
coordinates the flat metric can be written in the form $ds^2 = dzdz$
The infinitesimal generators of such transformations  close a Lie
algebra  $([G_n,G_m] = (n - m)G_{n+m})$.
\\
 In 2D CFT the stress energy tensor has only two components ,i.e., one analytic and the
other anti-analytic:
\begin{eqnarray}
\nonumber\ T = T_{zz}\\
\bar{T} = \bar{T}_{\bar{zz}}\\
\nonumber\ \partial_{z} \bar{T} = 0\\
\partial_{\bar{z}} T = 0
\end{eqnarray}
We can therefore expand for example $ T $ in Laurent series:
\begin{eqnarray}
T (z) = \sum^{\infty}_{ -\infty} \frac{L_{n}}{z^{n + 2}}.
\end{eqnarray}
In the quantum case the $ L_{n} $ become operators and relatively to
the commutator form a Virasoro algebra:
\begin{eqnarray}
[ L_{n} , L_{m}] = (n -m) L_{n + m} + \frac{C}{12} (m^{3} -
m)\delta_{m, -n}.
\end{eqnarray}
This is a central extension of the algebra $([G_n,G_m] = (n -
m)G_{n+m})$. The central extension in the commutator algebra arises
because of the normal ordering of the creation operators. It is a
standard result for quantum CFT in 2-D that the asymptotic density
of states for given $ L_{0} $ is completely determined by the
Virasoro algebra by means of the Cardy formula\cite{21}
\begin{eqnarray}
\rho (L_{0}) = \exp (2\pi \sqrt{\frac{c L_{0}}{6}}).
\end{eqnarray}
The entropy can therefore be calculated by using the logarithm of
the Cardy formula $ S = \ln \rho.$
\\
A central extension of the conformal algebra $([G_n,G_m] = (n -
m)G_{n+m})$ can already arise at classical level in the Poisson
algebra of the charges, as for example in the Liouville
theory\cite{22}.

\section{vii.Aproximate solution of the classical Liouville using Gaussion phase paket dynamics}
Their approach\cite{23},to classical Liouville equation draws on the
formal equivalence of this equation to the time dependent $
Schr\ddot{o}dinger $ equation\cite{24}.Gaussian wave packets,first
popularized by Heller\cite{25,26},for integrating the time dependent
$ Schr\ddot{o}dinger $ equation, have also been applied to the
calculation of specteroscopic correlation functions using the von
Neumann ( quantum Lioville ) equation\cite{27,28,29,30}.
\\
Mukamel and co-workers observed that the equation of motion for a
Gaussian approximation to the density matrix can be adapted to
classical mechanics by replacing the quantum Liouville operator and
phase space density distribution\cite{31}.The classical density
distribution of each particle is represented by a single Gaussian
phase packet.The many-body density distribution is expressed in a
Hartree approximation as a product of distributions for each
particle\cite{32,33}.
\\
Equation of motion for the density distribution are derived for both
constant energy and costant temperature dynamics and exact results
are obtained for the time evolution in the free particle and
harmonic potentials.We demonstrate the general applicability of the
Gaussian phase packet dynamics method through two application.The
first involves the calculation of equilibrium thermodynamic averages
for a Lennard-Jones cluster and fluid using Gaussian phase packets.

\section{viii. Quantum-classical Liouville dynamics of proton and deuteron transfer
rates in a solvated hydrogen-bonded complex} They studied\cite{34}
an intermolecular proton transfer reaction in a bulk polar solvent
of the form $ AH - B\rightleftharpoons A^{ - } - H^{ + }B $. The
model under study, which was constructed by Azzouz and
Borgis\cite{35}, describes a hydrogen-bonded phenol $ ( A ) $
trimethylamine $ ( B ) $ complex dissolved in methyl chloride. The
proton transfer rate constant and kinetic isotope effect $ (KIE) $
have been computed for this model using a wide variety of
techniques\cite{36,37,38,39,40,41,42,43}. The specific forms of the
interaction potentials, parameter values used, and the remaining
details of the model can be found in\cite{36,44} and\cite{45}. In a
previous work, they calculated the proton transfer rate constant for
this model with the $ AB $ distance constrained at $ R_{AB} = 2.7
{\AA} $.

\subsection{A:QUANTUM-CLASSICAL RATE THEORY}
The rate constant calculations are based on an expression for the
time dependent rate coefficient of the proton transfer reaction $ A
\rightleftharpoons B $
\begin{eqnarray}
k_{AB} (t) = \frac{1}{n^{eq}_{A}} \sum_{\alpha} \sum_{\acute{\alpha}
\geq \alpha} (2 - \delta_{\acute{\alpha} \alpha})\int dX \times Re [
N^{\alpha \acute{\alpha}}_{B} (X ,t) W^{\acute{\alpha} \alpha}_{A} (
X , \frac{i \hbar \beta}{2})]
\end{eqnarray}
 which is written in terms of a partial Wigner
representation of the bath degrees of freedom and a representation
of the protonic degrees of freedom in adiabatic states\cite{46}.
Here the bath phase space variables(coordinates and momenta,
respectively) are denoted by $ X = (R,P),n^{eq}_{A} $ is the
equilibrium density of species $A$ and $ \beta = \frac{1}{k_{B} T} $
is the inverse temperature. In this partial Wigner transform
representation\cite{47},the Hamiltonian $ \hat{H} $ of the system is
$ \hat{H}_{W} = \frac{P^{2}}{2M} + \hat{h}(R) $, where $ \hat{h}(R)
$ is the protonic Hamiltonian in the field of fixed bath particles
with mass $M$. The adiabatic basis states $ \{|\alpha;R\rangle \}
$are the solutions of the eigenvalue problem $
\hat{h}(R)|\alpha;R\rangle = E_{\alpha}(R)|\alpha;R\rangle $, where
$ \{E_{ \alpha }\} $ are the adiabatic energies. In this expression
for the rate coefficient, $ N^{\alpha \acute{\alpha}}_{B} (X,t) $is
the time evolved species $B$(product)operator, while $
W^{\acute{\alpha} \alpha }_{A} (X,\frac{i\hbar \beta }{2}) $is the
spectral density function that contains all information on the
quantum equilibrium structure. The solvent polarization reaction
coordinate\cite{48,49}.
\section{ix.statistical mechanics of non-Hamiltonian systems}
The dynamics of Hamiltonian systems is characterized by conservation
of phase space volume under time evolution\cite{50}, and this
conservation of phase volume is a cornerstone of conventional
statistical mechanics\cite{51,52}. Invariance of phase space volume
under Hamiltonian time evolution is the content of Liouvilles
theorem for divergenceless flows\cite{50,53}. At a deeper level,
conservation of phase space volume is understood to be a consequence
of the existence of an invariant symplectic form in the phase space
of Hamiltonian systems, and application of geometric methods and
concepts from the theory of differentiable
manifolds\cite{54,55,56,57,58}is essential for a fundamental
description of classical Hamiltonian systems\cite{50,54,59,60,61}.
\\
Non-Hamiltonian dynamics, characterized by nonzero phase space
compressibility\cite{53,62,63,64,65,66,67,68,69,70,71} is relevant
when we consider the statistical mechanics of thermostatted systems
\cite{72,73,74,75}. Such systems arise in the simulation of
ensembles other than microcanonical\cite{76}, and in the treatment
of nonequilibrium steady states\cite{72,73,75,77}.
\\
 Various homogeneous thermostatting mechanisms have been introduced to remove
heat supplied by nonequilibrium mechanical and thermal
perturbations. Phase space volume is no longer necessarily
conserved, and for nonequilibrium steady states the phase space
probability distribution is found to collapse onto a fractal set of
dimensionality lower than in the equilibrium
case\cite{72,73,75,78,79}. This phenomenon indicates a lack of
smoothness of the invariant measure in phase space in nonequilibrium
steady states\cite{75,80,81}. The dynamical evolution of the phase
space distribution function for Hamiltonian systems is described by
the Liouville equation\cite{50,51}. The Hamiltonian equation is
often considered be a special case of a so-called generalized
Liouville equation (henceforth GLE) appropriate for systems with
compressible dynamics\cite{53,62,63,64,65,66,67,68,69,70,71,82},
although the equation for the time-evolution of the Jacobian
determinant in a general compressible flow given in Liouvilles
original paper\cite{53} is, in fact, equivalent to the
GLE\cite{66,68,70}. A number of
authors\cite{62,63,64,65,66,67,68,69,70,71} have treated the
statistical mechanics of non-Hamiltonian systems in terms of the
GLE, and all have derived the result that the rate of change of the
Gibbs entropy for non-Hamiltonian systems is the ensemble average of
the divergence of the dynamical vector field (phase space
compressibility). Steeb\cite{74,67,82} applied the theory of Lie
derivatives and differential forms to derive the GLE for both
time-independent and time-dependent vector fields. Some explicit
solutions to the Liouville equation were given, and the existence of
singular solutions for systems with limit cycles (attracting
periodic orbits) was noted\cite{67}. The important paper by
Ramshaw\cite{71} gave a covariant formulation of the Liouville
equation and of the entropy. It was noted that invariant measures
(volume elements) associated with zero entropy production rate in
non-Hamiltonian systems must be smooth stationary solutions of the
GLE. Some limitations of the description of nonequilibrium steady
states in terms of the GLE were discussed by Holian et al\cite{83}.
\\
Tuckerman et al\cite{84} (hereafter, TEA) have recently applied
geometric methods from the theory of differentiable
manifolds\cite{55,56,57,58}, in particular the concepts of
Riemannian geometry\cite{57}, to the classical statistical mechanics
of non-Hamiltonian systems\cite{85,86,87}. TEA have argued that,
through introduction of so-called metric factors, it is always
possible to define a smooth invariant phase space measure in
non-Hamiltonian systems, even for nonequilibrium stationary
states\cite{85,86}. Moreover, the Gibbs entropy of the associated
phase space distribution function is found to be constant in time,
just as for Hamiltonian systems. TEA also claim that previous
formulations of the GLE (for example, papers\cite{72,73}) are in
some way incorrect, incomplete, or at least
coordinatedependent\cite{85,86,87}. These claims have proved
controversial\cite{88,89,90,91,92,84,93,94,95}.
\\
By definition, a coordinate-free formulation in the language of
differential forms\cite{54,55,56,57,58} removes any
question\cite{87} concerning the coordinate dependence of any
results obtained. The apparatus of differential forms is the
appropriate machinery for treating the transformations of variables
and volume elements arising in the dynamics of both Hamiltonian and
non-Hamiltonian systems. For any region of phase space, the fraction
of the ensemble inside the region is obtained by integration of the
density n-form over the region. The density n-form can be written as
the product of a volume form\cite{55,56} (comoving volume element)
and a phase space distribution function. The GLE, which describes
the evolution of the phase space distribution function, then follows
from the transport equation\cite{67}.
\\
 To determine the fraction of
the ensemble in a given region, we simply need to count ensemble
members. As we do not need a volume form to count ensemble members
inside a prescribed region of phase space, the density n-form $ \rho
$ is defined without reference to any particular volume form. Any
result expressed in terms of $ \rho $ alone is therefore manifestly
independent of the choice of volume form on the phase space
manifold. This covariance with respect to the choice of volume form
is the essential advantage of a description of the ensemble density
in terms of the n-form  $ \rho $ .
\\
 When considering the phase space
structure and dynamics of Hamiltonian and non- Hamiltonian systems,
the notion of distance associated with the familiar properties of
Riemannian manifolds\cite{55} is irrelevant. Volume forms provide
exactly the construct needed, namely, a definition of volume without
distance. Moreover, the Lie derivative of the volume form provides a
definition of divergence without metric connection. The existence of
an invariant volume form is important for simulation of equilibrium
properties via non-Hamiltonian dynamics; if the dynamics preserves a
given volume form (invariant measure) then, provided the system is
ergodic, and that all relevant constraints are taken into
account\cite{87}, phase space averages with respect to the invariant
measure can be evaluated by computing long-time averages over a
single trajectory.For all the equilibrium non- Hamiltonian systems
discussed to date, a smooth stationary invariant measure can be
found\cite{87}. Nevertheless, any system with net attracting or
repelling periodic orbits cannot possess a smooth invariant
measure\cite{96}, so that, contrary to the assumption made
in\cite{86}, smooth invariant measures for arbitrary non-Hamiltonian
systems do not exist.True invariant measures, which are in general
singular (not absolutely continuous with respect to the usual volume
element), can be defined in terms of infinite-time averages of the
density n-form $ \rho_{t} $ \cite{81}.
\\
If the metric tensor is stationary, then compatibility of the
Riemannian structure with the dynamics requires the associated
metric factor $ \sqrt{g} $ to be a time-independent solution of the
GLE. In this case, the metric factor defines an invariant
(Riemannian) volume form in the usual fashion\cite{55}. On the other
hand, if the metric tensor is allowed to be
time-dependent\cite{87,97,98}, then compatibility requires the
associated metric factor to be a non-stationary solution of the
Liouville equation. The entropy defined with respect to the
associated time-dependent Riemannian volume form is then
constant\cite{87}; for nonequilibrium steady states, the underlying
metric factor will become ever more nearly singular (fractal) at
long times, so that this result is only of formal
significance\cite{99}.
\section{x:small-time parameter method in Liouville equation}
Considering the formal asymptotic solution of Cauchy problem $[
\frac{\partial}{\partial t}f(q,p,t) = Lf(q,p,t) , f(q,p,t)|_{t=0} =
f_{0} (q,p)] $ by the method of the small time space in\cite{100}
and defined by mapping $ \tau = 1 - \exp (- st) , s > 0 $ Gennady
Rudykh , Alexander Sinitsyn and Eugene Dulov ,transformed the
initial infinite time interval $ R^{+} $ onto a small finite one $ J
\triangleq \{ \tau : 0\leq \tau < 1 \} . $
\\
This technique is quite general and was applied to Liouville
equation by G.Rudykh and A.Sinitsyn \cite{101} at the earlier 80's
of the 20th century.
\\
The small-time transformation is also well known to the applied
mathematicians because it suits for numerical integration over the
semi-infinite intervals. The obvious benefit for a such kind of
transform lies in power series expansion over a time scale.Using
finite interval $ [0,1] $ one can pay the attention to the series
coefficients and their convergence properties.
\\
In our particular case a transformed Cauchy problem becomes:
\begin{eqnarray}
(1-\tau)\frac{\partial}{\partial \tau} f(q,p,\tau) = \frac{1}{s} L
f(q,p,\tau)
\\
 \nonumber f(q,p,\tau) |_{\tau = 0} = f_{0}(q,p)
 \end{eqnarray}
A solution of the Cauchy problem $ (75) $ in the small-time space
was studied in the form of asymptotic expansion:
\begin{eqnarray}
f(q,p,\tau) = \sum^{\infty}_{k=0} f_{k}(q,p)\cdot \tau^{k}.
\end{eqnarray}
Substituting $ (76) $ in to $ (75) $ and equating the coefficients
for $\tau $ , one obtains\cite{102}
\begin{eqnarray}
f_{k}(q,p) = \frac{k - 1}{k} f_{k-1}(q,p) + \frac{1}{s k}(
[H(q,p),f_{k-1}(q,p)] - \sum^{n}_{i=1} \frac{\partial}{\partial
p_{i}} \{Q^{*}_{i}(q,p)\cdot f_{k-1}(q,p)\})
\end{eqnarray}
Hence
\begin{eqnarray}
f_{k}(q,p) = \frac{1}{s k} L [\prod^{k -1}_{r=1}(1 + \frac{1}{s
r}L)] f_{0}(q,p)
\\
\nonumber k = 2,3,...
\end{eqnarray}
With
\begin{eqnarray}
\prod^{k-1}_{r=1} (1+\frac{1}{s k} L) = a_{k-1} + \frac{1}{s}
a_{k-2} L + ... + \frac{1}{s^{k-1}} a_{0} L^{k-1}
\end{eqnarray}
Where $ a_{0} = \frac{1}{(k-1)!} , a_{k-1} = 1 .$
\section{appendix}
In this short section we present a concise  proof for the theorem
2.A
\\
2.A:The family $S^{N}_{\Lambda}(t),t\in ]-\infty,+\infty[
$,constitutes a strongly continuous one -operator group of isometric
operators in $L_{N}(\Lambda)$ whose infinitesimal operator
$H^{N}_{\Lambda}$ is given on $L^{0}_{N}(\Lambda)$ by the $Poisson$
bracket with the Hamiltonian of system of non-interacting particles
with bondary conditions.
\\
PROOF.Group properties of the family $S^{N}_{\Lambda}(t)$ follow
from the group property of  a phase trajectory $X(t,\chi)$. For
$f_{N} \in L^{0}_{N}(\Lambda)$ we show that:
\begin{eqnarray}
\lim_{\Delta t\rightarrow o}\| S^{N}_{\Lambda}(t + \Delta t)f_{N} -
S^{N}_{\Lambda}(t)f_{N} \| = 0.
\end{eqnarray}
Indeed,for $f_{N} \in
L^{0}_{N}(\Lambda)$ since $S^{N}_{\Lambda}(t)$ is isometric we have
\begin{eqnarray}
\nonumber \|S^{N}_{\Lambda}(t + \Delta t)f_{N} -
S^{N}_{\Lambda}(t)f_{N}\| = \|S^{N}_{\Lambda}(\Delta t)f_{N} -
f_{N}\| = \int dx |f_{N}(X(\Delta t,\chi)) - f_{N}(\chi)|.
\end{eqnarray}
Since for sufficiently small $\Delta t$ and $f_{N}(X(\Delta
t,\chi))$ is non-zero outside some neighborhood of forbidden
configurations and the boundary $\partial \Lambda$ of the domain
$\Lambda$,where there are no particle collisions,we have
$f_{N}(X(\Delta t,\chi)) \rightarrow f_{N}(\chi)$ uniformly with
respect to $\chi$ as $\Delta t \rightarrow 0.$Thus we can carry out
the $\lim \Delta t \rightarrow 0$ under the integral,hence the
validity of (79) is proved.Since $L^{0}_{N}(\Lambda)$ is dense
everywhere in $L_{N}(\Lambda),$from (79) and the boundedness of the
group $S^{N}_{\Lambda}(t)$ there follows the strong continuty of the
group $S^{N}_{\Lambda}(t)$ in $L_{N}(\Lambda).$


\begin{thebibliography}{35}
\bibitem{1}
Debra.J.Searles and Denis.J.Evans,The Fluctuation Theorem and
Green-KUbo Relations,arxive:cond-matt/[9902021]
\\
\bibitem{2}
   M.V.Berry,True Quantum Chaos? An Instructive Example.,Springer
   proceedings in physics, Vol.58,New Tradns in Nuclear Collective
   Dynamics,Eds: Y.Abe.H.Horiuchi,K.Matsuyangi,Springer-Veriag
   Berlin Heidelberg 1991.
\\
\bibitem{3}
D.Ya.Petrina and V.I.Gerasimenko,Mathematical problems of statistical
mechanics of a system of elastic balls,Uspekhi Mat.Nauk 45:3 (1990),
135-182,Russuian Math.surveys 45:3 (1990),153-211
\\
\bibitem{4}
Zhenglu Jiang,On the Liouville equation, Transport theory and
statistical physics journal,Vol.31,No.3,pp.267 272,2002
\\
\bibitem{5}
P.Avakian,V.Ern,R.E.Merrifield,and A.Suna,Phys.Rev.165,974(1960).
\\
\bibitem{6}
P.Reineker,in Exction Dynamics in Molecular Crystals and Aggregates,
edited by G.Hõhler ( Springer,Berlin,1982).
\\
\bibitem{7}
V.M.Kenker,in Excition Dynamics in Molecular Crystals and
Aggregates,edited by G.Hõhler ( Springer,Berlin,1982).
\\
\bibitem{8}
Y.Kagan and M.I.Klinger,J.Phys.C7,2791 (1974).
\\
\bibitem{9}
D.W.Brown and V.M.Kenker,in Electronic Structure and Properties of
Hydrogen in Metals,Vol.6 of NATO Conference Series \emph{VI}:
Materials Science, editted by P.Jena and C.B.Satterthwaite
(Plenum,New York,1983),P.177.
\\
\bibitem{10}
V.M.Kenker and D.W.Brown,Exact solution of the stochastic Liouville
equation and application to an evaluation of the neutron scattering
function,Phys.Rev.B.Vol.31,Number 4 (1985).
\\
\bibitem{11}
V.M.Kenker,Phys.Rev.B 22,2089 (1980);Z.Phys.B 43,22 (1981).
\\
\bibitem{12}
Alexander J.Vega and Daniel Fiat,The Stochastic Liouville Equation and
The Approach to Thermal Equilibrium,Department of structural Chemstry,
The Weizmann Institute of Science,Rehovot,Israel
\\
\bibitem{13}
R.Kubo,J.Phys.Soc.Japan,26,Suppl,1(1969).
\\
\bibitem{14}
N.Bloembergen,E.M.Purcell and R.V.Pound,Phys.Rev.73,679(1948).
\\
\bibitem{15}
R.K.Wangness and F.Bloch,Phys.Rev.89,728(1953).
\\
\bibitem{16}
F.Bloch,Phys.Rev.102,104(1956).
\\
\bibitem{17}
F.Bloch,Phys.Rev.105,1206(1957).
\\
\bibitem{18}
R.Kubo,Adv.Chem.Phys.15,101(1969).
\\
\bibitem{19}
See,e.g.,E.W.Montroll and B.West,in Fluctuation Phenomena,edited by
E.W.Montroll and J.L.Lebowitz
(North-Holland,Amesterdam,1979);J.Stat.Phys.13,17(1975),and
references therein;E.W.Montroll,Energetics in Metallurgical
Phenomena (Gordon and Breach,New York,1967),Vol.3.
\\
\bibitem{20}
Alex GIACOMINIl,Proceedings of Institute of Mathematics of NAS of
Ukraine 2004, Vol. 50, Part 2, 767–773
\\
\bibitem{21}
Cardy J.A., Operator content of two-dimensional conformally
invariant theories, Nuc. Phys. B, 1986, V.270,186 to 204.
\\
\bibitem{22}
Jackiw R., Liouville field theory: a two-dimensional model for
gravity, In Quantum Theory of Gravity,Essays in Honour of B. De
Witt, Editor S. Christensen, Bristol, Hilger, 1984, 403 to 420.
\\
\bibitem{23}
Jianpeng Ma,D.Hsu and John E.Straub,Approximate solution of the
classical Liouville equation using Gaussian phase packet
dynamics:Application to enhanced equilibrium averaging and global
optimization,J.chem.phys,Vol.99,No.5,1 september 1993.
\\
\bibitem{24}
B.J.Berne and R.Pecora,Dynamic Light Scattering ( Wiley
Interscience,New york,1976).
\\
\bibitem{25}
E.J.Heller,J.chem.phys.62,1544(1975);Acc.chem.Rcc.14,368(1981).
\\
\bibitem{26}
D.Thirumalai,E.Bruskin and B.J.Berne,J.chem.phys.83,230(1985).
\\
\bibitem{27}
R.D.Coalson and
M.Karplus,J.chem.phys.79,6150(1983);81,2891(1984);93,3919(1990);see
also
A.DMcLachlan,Mol.phys.8939(1964);E.J.Heller,J.chem.phys.64,63(1976).
\\
\bibitem{28}
S.Mukamel,J.chem.phys.88,3185(1984).
\\
\bibitem{29}
J.Grad,Y.J.Yan,A.Haque and
S.Mukamel,chem.phys.Lett.134,219(1987);J.chem.phys.86,3441(1987).
\\
\bibitem{30}
Y.J.Yan,and,S.Mukamel,J.chem.phys.88,5735(1988);S.Mukamel,and
Y.J.Yan,Adv.chem.phys.73,579(1989).
\\
\bibitem{31}
J.Grad,Y.J.Yan,A.Haque and
S.Mukamel,chem.phys.Lett.134,219(1987);J.chem.phys.86,3441(1987).
\\
\bibitem{32}
R.Elber,and,M.Karplus,J.Am.chem.soc.112,9161(1990);R.Czerminski,and,R.Elber,proteins.10,70(1991).
\\
\bibitem{33}
R.B.Gerber,and,M.A.Ratner.Adv.chem.phys.70.79(1988);G.C.Schatz,V.Buch,M.A.Ratner
and,R.B.Gerber,J.chem.phys.79.1808(1983);V.Buch,R.B.Gerber,and
M.A.Ratner,chem.phys.Lett.101,44(1983);R.B.Gerber,V.Buch,and,
M.A.Ratner,J.chem.phys.77,3022(1982);chem.phys.Lett.91,173(1982).
\\
\bibitem{34}
Gabriel Hanna and Raymond Kapral,Quantum-classical Liouville
dynamics of proton and deuteron transfer rates in a solvated
hydrogen-bonded complex,DOI: 10.1063/1.2907847,2008 American
Institute of Physics.
\\
\bibitem{35}
H. Azzouz and D. Borgis, J. Chem. Phys. 98, 7361 (1993).
\\
\bibitem{36}
S. Hammes-Schiffer and J. C. Tully, J. Chem. Phys. 101, 4657 (1994).
\\
\bibitem{37}
H. Azzouz and D. Borgis, J. Chem. Phys. 98, 7361 (1993).
\\
\bibitem{38}
A. Staib, D. Borgis, and J. T. Hynes, J. Chem. Phys. 102, 2487
(1995).
\\
\bibitem{39}
D. Antoniou and S. D. Schwartz, J. Chem. Phys. 110, 465 (1999).
\\
\bibitem{40}
D. Antoniou and S. D. Schwartz, J. Chem. Phys. 110, 7359 (1999).
\\
\bibitem{41}
R. P. McRae, G. K. Schenter, B. C. Garrett, Z. Svetlicic, and D. G.
Truhlar, J. Chem. Phys. 115, 8460 (2001).
\\
\bibitem{42}
S. Y. Kim and S. Hammes-Schiffer, J. Chem. Phys. 119, 4389 (2003).
\\
\bibitem{43}
T. Yamamoto and W. H. Miller, J. Chem. Phys. 122, 044106 (2005).
\\
\bibitem{44}
G. Hanna and R. Kapral, J. Chem. Phys. 122, 244505 (2005).
\\
\bibitem{45}
G. Hanna and R. Kapral, Acc. Chem. Res. 39, 21 (2006).
\\
\bibitem{46}
H. Kim and R. Kapral, J. Chem. Phys. 123, 194108 (2005).
\\
\bibitem{47}
E. Wigner, Phys. Rev. 40, 749 (1932).
\\
\bibitem{48}
R. A. Marcus and N. Sutin, Biochim. Biophys. Acta 811, 265 (1985).
\\
\bibitem{49}
A. Warshel, J. Phys. Chem. 86, 2218 (1982).
\\
\bibitem{50}
V.I. Arnold, Mathematical Methods of Classical Mechanics (Springer,
New York, 1978).
\\
\bibitem{51}
J.W. Gibbs, Elementary Principles in Statistical Mechanics (Yale
University Press, New Haven, 1902).
\\
\bibitem{52}
R.C. Tolman, The Principles of Statistical Mechanics (Oxford
University Press, Oxford, 1938).
\\
\bibitem{53}
J. Liouville, Sur la Théorie de la Variation des constantes
arbitraires, J. Math. Pures Appl. 3 (1838) 342–349.
\\
\bibitem{54}
R. Abraham and J.E. Marsden, Foundations of Mechanics
(Benjamin/Cummings, Reading, MA, 1978).
\\
\bibitem{55}
B. Schutz, Geometrical Methods of Mathematical Physics (Cambridge
University Press, Cambridge, 1980).
\\
\bibitem{56}
R. Abraham, J.E.Marsden and T. Ratiu, Manifolds, Tensor Analysis,
and Applications (Springer, New York, 1988).
\\
\bibitem{57}
B.A Dubrovin, A.T. Fomenko and S.P. Novikov, Modern Geometry –
Methods and Applications, Part 1. The Geometry of Surfaces,
Transformation Groups, and Fields (Springer, New York, 1992).
\\
\bibitem{58}
T. Frankel, The Geometry of Physics (Cambridge University Press,
Cambridge, 1997).
\\
\bibitem{59}
J.V. José and E.J. Saletan, Classical Dynamics (Cambridge University
Press, Cambridge, 1998).
\\
\bibitem{60}
J.E. Marsden and T.S. Ratiu, Introduction to Mechanics and Symmetry
(Springer, New York, 1999).
\\
\bibitem{61}
S.F. Singer, Symmetry in Mechanics (Birkhäuser, New York, 2001).
\\
\bibitem{62}
S. Guiasu, La méchanique statistique non conservative, Rev. Roum.
Math. Pures Appl. 11 (1966) 541–557.
\\
\bibitem{63}
G. Gerlich, Die Verallgemeinerte Liouville-Gleichung, Physica 69
(1973) 458–466.
\\
\bibitem{64}
W.-H. Steeb, Generalized Liouville equation, entropy and dynamic
systems containing limit cycles, Physica A 95 (1979) 181–190.
\\
\bibitem{65}
J. Fronteau, Vers une description non conservative de l'évolution en
physique, Hadronic J. 2 (1979) 727–829.
\\
\bibitem{66}
M. Grmela, J. Fronteau and A. Tellez-Arenas, Inverse Liouville
problem, Hadronic J. 3 (1980) 1209– 1241.
\\
\bibitem{67}
W.-H. Steeb, A comment on the generalized Liouville equation, Found.
Phys. 10 (1980) 485–493.
\\
\bibitem{68}
J. Fronteau, Liouville's theorem as a link between different
viewpoints, Hadronic J. 5 (1982) 577–592.
\\
\bibitem{69}
L. Andrey, The rate of entropy change in non-Hamiltonian systems,
Phys. Lett. A 111 (1985) 45–46.
\\
\bibitem{70}
L. Andrey, Note concerning the paper "The rate of entropy change in
non-Hamiltonian systems", Phys. Lett. A 114 (1986) 183–184.
\\
\bibitem{71}
J.D. Ramshaw, Remarks on entropy and irreversibility in
non-Hamiltonian systems, Phys. Lett. A 116 (1986) 110–114.
\\
\bibitem{72}
D.J. Evans and G.P. Morriss, Statistical Mechanics of Nonequilibrium
Liquids (Academic, New York, 1990).
\\
\bibitem{73}
W.G. Hoover, Computational Statistical Mechanics (Elsevier, New
York, 1991).
\\
\bibitem{74}
G.P. Morriss and C.P. Dettmann, Thermostats: Analysis and
application, Chaos 8 (1998) 321–336.
\\
\bibitem{75}
J.R. Dorfman, An Introduction to Chaos in Nonequilibrium Statistical
Mechanics (Cambridge University Press, Cambridge, 1999).
\\
\bibitem{76}
S. Nosé, Constant temperature molecular dynamics methods, Prog.
Theor. Phys. Suppl. 103 (1991) 1–46.
\\
\bibitem{77}
C.J. Mundy, S. Balasubramanian, K. Bagchi, M.E. Tuckerman, G.J.
Martyna and M.L. Klein, Nonequilibrium Molecular Dynamics, Reviews
in Computational Chemistry, Vol. 14 (Wiley/VCH, New York, 2000) pp.
291–397.
\\
\bibitem{78}
B.L. Holian, W.G. Hoover and H.A. Posch, Resolution of Loschmidt's
paradox: The origin of irreversible behavior in reversible atomic
dynamics, Phys. Rev. Lett. 59 (1987) 10–13.
\\
\bibitem{79}
W.G. Hoover, H.A. Posch, K. Aoki and D. Kusnezov, Remarks on
non-Hamiltonian statistical mechanics: Lyapunov exponents and phase
space dimensionality loss, Europhys. Lett. 60 (2002) 337– 341.
\\
\bibitem{80}
J.-P. Eckmann and D. Ruelle, Ergodic theory of chaos and strange
attractors, Rev. Mod. Phys 57 (1985) 617–656.
\\
\bibitem{81}
D. Ruelle, Smooth dynamics and new theoretical ideas in
nonequilibrium statistical mechanics, J. Stat. Phys. 95 (1999)
393–468.
\\
\bibitem{82}
W.-H. Steeb, The Lie derivative, invariance conditions and physical
laws, Z. Naturforsch. A 33 (1978) 742–748.
\\
\bibitem{83}
B.L. Holian, G. Ciccotti, W.G. Hoover, B. Moran and H.A. Posch,
Nonlinear-response theory for time-independent fields: Consequences
of the fractal nonequilibrium distribution function, Phys. Rev. A 39
(1989) 5414–5421.
\\
\bibitem{84}
M.E. Tuckerman, C.J. Mundy, S. Balasubramanian and M.L. Klein,
Response to "Comment on 'Modified nonequilibrium molecular dynamics
for fluid flows with energy conservation' [J. Chem. Phys. 108 (1998)
4351]", J. Chem. Phys. 108 (1998) 4353–4354.
\\
\bibitem{85}
M.E. Tuckerman, C.J. Mundy and M.L. Klein, Toward a statistical
thermodynamics of steady states, Phys. Rev. Lett. 78 (1997)
2042–2045.
\\
\bibitem{86}
M.E. Tuckerman, C.J. Mundy and G.J. Martyna, On the classical
statistical mechanics of non- Hamiltonian systems, Europhys. Lett.
45 (1999) 149–155.
\\
\bibitem{87}
M.E. Tuckerman, Y. Liu, G. Ciccotti and G.J. Martyna,
Non-Hamiltonian molecular dynamics: Generalizing Hamiltonian phase
space principles to non-Hamiltonian systems, J. Chem. Phys. 115
(2001) 1678–1702.
\\
\bibitem{88}
P. Reimann, Comment on "Toward a statistical thermodynamics of
steady states", Phys. Rev. Lett. 80 (1998) 4104.
\\
\bibitem{89}
W.G. Hoover, D.J. Evans, H.A. Posch, B.L. Holian and G.P. Morriss,
Comment on "Toward a statistical thermodynamics of steady states",
Phys. Rev. Lett. 80 (1998) 4103.
\\
\bibitem{90}
M.E. Tuckerman, C.J. Mundy and M.L. Klein, Reply to comment on
"Toward a statistical thermodynamics of steady states", Phys. Rev.
Lett. 80 (1998) 4105–4106.
\\
\bibitem{91}
W.G. Hoover, Liouville's theorems, Gibbs' entropy, and multifractal
distributions for nonequilibrium steady states, J. Chem. Phys. 109
(1998) 4164–4170.
\\
\bibitem{92}
D.J. Evans, D.J. Searles, W.G. Hoover, C.G. Hoover, B.L. Holian,
H.A. Posch and G.P. Morriss, Comment on "Modified nonequilibrium
dynamics for fluid flows with energy conservation [J. Chem. Phys.
106 (1997) 5615]", J. Chem. Phys. 108 (1998) 4351–4352.
\\
\bibitem{93}
W.-J. Tzeng and C.-C. Chen, The statistical thermodynamics of steady
states, Phys. Lett. A 246 (1998) 52–54.
\\
\bibitem{94}
W.G. Hoover, The statistical thermodynamics of steady states, Phys.
Lett. A 255 (1999) 37–41.
\\
\bibitem{95}
J.D. Ramshaw, Remarks on non-Hamiltonian statistical mechanics,
Europhys. Lett. 59 (2002) 319– 323.
\\
\bibitem{96}
I.P. Cornfeld, S.V. Fomin and Y.G. Sinai, Ergodic Theory (Springer,
New York, 1982).
\\
\bibitem{97}
G. Sardanashvily, The Lyapunov stability of first order dynamics
equations with respect to timedependent Riemannian metrics, Preprint
(2002).
\\
\bibitem{98}
G. Sardanashvily, The Lyapunov stability of first order dynamics
equations with respect to timedependent Riemannian metrics. An
example, Preprint (2002).
\\
\bibitem{99}
Gregory S. Ezra,On the statistical mechanics of non-Hamiltonian
systems: the generalized Liouville equation, entropy, and
time-dependent metrics,Journal of Mathematical Chemistry Vol. 35,
No. 1, January 2004
\\
\bibitem{100}
A.N.PANCHENKOV,The foundations of the theory of limit
well-posesness,(In Russia), Nauka,Moscow,(1976),240p.
\\
\bibitem{101}
A.V.Sinitsyn,Asymptotic solution of the Cauchy problem for the
generalized Liouville equation, In: "Perturbation Methods in
Mechanics" , (In Russia) , Irkutsk Computing Center of SB of USSR,
Irkutsk, (1984) ,pp.180-187.
\\
\bibitem{102}
G.A.Rudykh, Asymptotic solution of the Liouville equation for
nonconservative systems of forces ,In: Asymptotic Methods in theory
of systems,(In Russia), Irkutsk Energetic Institute of SB of
USSR,Irkutsk,(1978),pp.71-86.
\\
\bibitem{103}
T.Kato,Perturbation theory for linear operators,springer,New York 1966.
Translation: Teoriya vazmushchenii lineinykh operator,Mir,Moscow 1972.
\\
\bibitem{104}
D.Ya.Petrina,V.I.Gersimenko, and P.V.Malyshev,Matematicheskie osnovy
klassicheskoi statusticheskoi mekhaniki (Mathematical foundations of
classical statistical mechanics),Naukova Dumka,Kiev 1985. MR 88
e:82009.
\\
\bibitem{105}
V.P.Maslov and S.E.Tariverdiev,The asymptotic behaviour of the
Kolmogorov Feller equation for a system of a large number of
particles, Probablity Theory. Mathematical Statistics.Theoretical
Cybernetics,Vol.19,VINITI,Moscow 1982,pp.85-126. MR 84 j:82064
\\



\end{thebibliography}
\end{document}